\documentclass[12pt]{article}
\usepackage{graphics}
\usepackage{latexsym, pstricks}
\usepackage{amsmath,amssymb,amsthm}

\font\titulo=cmbx10 scaled\magstep1 
scaled\magstep1

\def\section#1{\vskip 1.5truepc plus 0.1truepc minus 0.1truepc
    \goodbreak \leftline{\titulo#1} \nobreak \vskip 0.1truepc
    \indent}
\def\frc#1#2{\leavevmode\kern.1em
    \raise.5ex\hbox{\the\scriptfont0 $ #1 $}\kern-.1em
    /\kern-.15em\lower.25ex\hbox{\the\scriptfont0 $ #2 $}}

\def\IZ{{\rm Z}\llap{\vrule height7.1pt width1pt
     depth-.4pt\phantom t}} 

\newbox\pmbbox
 \def\pmb#1{{\setbox\pmbbox=\hbox{$#1$}%
\copy\pmbbox\kern-\wd\pmbbox\kern.3pt\raise.3pt\copy\pmbbox\kern-\wd\pmbbox
\kern.3pt\box\pmbbox}}

\font\cmss=cmss10 \font\cmsss=cmss10 at 7pt

\def\IZ{\relax\ifmmode\mathchoice
{\hbox{\cmss Z\kern-.4em Z}}{\hbox{\cmss Z\kern-.4em Z}}
{\lower.9pt\hbox{\cmsss Z\kern-.4em Z}} {\lower1.2pt\hbox{\cmsss
Z\kern-.4em Z}}\else{\cmss Z\kern-.4em Z}\fi}

\font\cmss=cmss10 \font\cmsss=cmss10 at 7pt
\def\IS{\relax\ifmmode\mathchoice
{\hbox{\cmss S\kern-.4em S}}{\hbox{\cmss S\kern-.4em S}}
{\lower.9pt\hbox{\cmsss S\kern-.4em S}} {\lower1.2pt\hbox{\cmsss
S\kern-.4em S}}\else{\cmss S\kern-.4em S}\fi}

\begin{document}

\centerline{\titulo $ \mathbb{Z}_2 $-Algebras in the Boolean}

\centerline{\titulo Function Irreducible Decomposition}

\vskip 1.2pc \centerline{Martha Takane \ and \ Federico Zertuche}

\vskip 1.2pc \centerline{Instituto de Matem\'aticas, Unidad
Cuernavaca} \centerline{Universidad Nacional Aut\'onoma de
M\'exico} \centerline{A.P. 273-3, 62251 Cuernavaca, Mor.,
M\'exico.} \centerline{\tt takane@matcuer.unam.mx \ \
zertuche@matcuer.unam.mx}

\vskip 3pc {\bf \centerline {Abstract}}

We develop further the consequences of the irreducible-Boolean classification established in Ref.~[9]; which have the advantage of allowing strong statistical calculations in disordered Boolean function models, such as the \textit{NK}-Kauffman networks. We construct a ring-isomorphism $ \mathfrak{R}_K \left\{ i_1, \dots, i_\lambda \right\} \cong \mathcal{P}^2 \left[ K \right] $ of the set of reducible $K$-Boolean functions that are reducible in the Boolean arguments with indexes $ \left\{ i_1, \dots, i_\lambda \right\} $; and the double power set $ \mathcal{P}^2 \left[ K \right] $, of the first $K$ natural numbers. This allows us, among other things, to calculate the number $ \varrho_K \left( \lambda, \omega \right) $ of $K$-Boolean functions which are $ \lambda $-irreducible with weight $ \omega $.  $ \varrho_K \left( \lambda, \omega \right) $ is a fundamental quantity in the study of the stability of \textit{NK}-Kauffman networks against changes in their connections between their Boolean functions; as well as in the mean field study of their dynamics when Boolean irreducibility is taken into account.

\vskip 2pc

\noindent {\bf Short title:} {\it $ \mathbb{Z}_2 $-Algebras and Irreducibility}



\newpage

\baselineskip = 14.8pt

\section{1. Introduction}

Boolean functions play a seminal role in pure, and in applied mathematics. In pure mathematics: they constitute a field of study as themselves because of their rich structure. They appear in the theory of functional-graphs, as well as in the general theory of graphs, just to quote some related fields~${}^{1}$. In applied mathematics: they support all the architecture of the modern digital computers as well as many fields of theoretical physics, theoretical biology, statistical mechanics, etc~${}^{2-5}$.

Our particular interest in them is mainly based in the work of Stuart Kauffman which introduced the (now so-called) \textit{NK}-Kauffman networks in order to try to understand how biological processes evolve; in an apparently spontaneously way, from disorder to order~${}^{3,6}$. \textit{NK}-Kauffman networks are an example of disordered systems of many variables (Boolean functions) with deterministic rules of evolution. They are constructed in a random way, that trays to mimic the way in which nature constructs the basic bricks of the animated world; allowing a statistical treatment for them. One of the main problems with \textit{NK}-Kauffman networks is to understand the way their dynamics changes as the fundamental parameters of the model change {\it i.e.}: the number of Boolean functions $N$, its mean connectivity $K$, and its probability bias $p$ that a Boolean function gives as output a ``1" for a preassigned input~${}^{7-11}$. This, however has only been done in an exact statistical way (including asymptotic expansions), only for special values of $ N $, $ K $, and $ p $: The case of the so-called {\it random map model} where $ K = N $, and $ p = 1/2 $ (so the Boolean functions are extracted with equiprobability)~${}^{7}$. And the case $ K = 1 $, also with $ p = 1/2 $~${}^{8}$.

In Ref.~[9], a new classification of Boolean functions, in terms of their real connectivity, was established as a necessity to understand new aspects of their dynamics as a statistical system. This led to the concept of {\it irreducible-degree} of a Boolean function, and allowed to understand new asymptotic properties of \textit{NK}-Kauffman networks~${}^{9-11}$. We stress that this classification should be not confused with the one developed by Kauffman, which divides Boolean functions into {\it canalizing} and {\it no-canalizing}~${}^{3,12}$. While canalizing classification is a first step to understand the dynamics of \textit{NK}-Kauffman networks; it gives less information about a Boolean function, than its irreducible-degree. As examples: {\it i}) The injectivity properties of the genotype-phenotype function $ \Psi $ which is responsible of the dynamical diversity in \textit{NK}-Kauffman networks, are obtained by an expansion in terms of the irreducible-degree~${}^{9,11}$. {\it ii}) The calculation for the probability that an \textit{NK}-Kauffman network remains invariant against a change in one of their $K$-connection functions depends on a series that involves term by term the irreducible-degree of the Boolean functions~${}^{9}$. See also Eq.~(A1). {\it iii}) The dynamics of \textit{NK}-Kauffman networks study through the mean field approach is based in their mean connectivity, which is regulated by the irreducible-degree, and not by the canalizing classification. See Appendix~A for a detailed discussion about canalization.

While our specific research is based in Kauffman's networks, the need of new mathematical tools to deal with them has shove us to study the relation of Boolean functions with rings. Thus, in this work we construct rings for the sets of Boolean functions which are reducible in some of their arguments and use this powerful isomorphism to calculate the number of Boolean functions with irreducible-degree $ \lambda $, and weight $ \omega $, denoted by $ \varrho_K \left( \lambda, \omega \right) $. We also established an analytic expression for the irreducible-degree $\lambda $ of a given Boolean function.

The content of this article is as follows: In Sec.~2, we introduce the formal apparatus, defining the concept of irreducible-degree of a Boolean function originally proposed in Ref.~[9]. We establish their classification in terms of it, and some previous developed counting formulas are settled. It is also introduced the concept of weight, and some useful decompositions of the space of Boolean functions are shown. In Sec.~3, we recast the terminology of Boolean functions into the language of set theory by associating to each Boolean function on $K$ arguments a set $ \mathcal{B} $ which is an element of $ {\cal P}^2 \left[ K \right] $, the double power set of the first $K$ natural numbers. With this framework we further study the mathematical structure of the Boolean functions. Sec.~4, contains our main results; there we find: {\it i}) An analytical expression for the irreducible-degree $ \lambda \left( b_K \right) $ of a Boolean function $ b_K $ on $K$ arguments. \break {\it ii}) A subring of the set of reducible functions on $ \lambda $ arguments, of the ring ($ \mathcal{P}^2 \left[ K \right] $, $ \triangle $, $ \cap $). {\it iii}) Last but not least, we find an analytic expression for $ \varrho \left( \lambda, \omega \right) $, the number of Boolean functions $ b_K $ with irreducible-degree $ \lambda $ and weight $ \omega $. In Sec.~5 we discuss our results. In Appendix~A, the mathematical definition of a canalizing function is introduced, and the difference with their irreducible-degree is explained in detail. In Appendix~B we make a summary of the treatment developed in Ref.~[10], where a study of the phase transition curve of \textit{NK}-Kauffman networks is done.

\bigskip

\section{2. Boolean Functions and their Irreducible Classification}

Along this work: We use the convention that the set of natural numbers starts with the number $ 1 $, {\it i.e.} $ \mathbb{N} = \left\{ 1, 2, 3, \dots \right\} $. And we denote by $ \left[ K \right] = \left\{ 1, 2, \dots, K \right\} $  the set of the first $ K $ natural numbers. For all $ S, \tilde{S} \in \mathbb{Z}_2 $, additions $ S + \tilde{S} $ are intended to be {\it modulo $2$}. When clarity will be required we emphasize this by $ \left[ S + \tilde{S} \right]_2 $.


\begin{itemize}

\item[] {\bf Definition}~1:

\item[{\it i})] A {\it $K$-Boolean function} is a map
$$
b_K: \mathbb{Z}_2^K \to \mathbb{Z}_2, \eqno(1)
$$
with the assignment
$$
{\bf S} = \left( S_1, \dots, S_K \right) \longmapsto \sigma . \eqno(2)
$$
\item[{\it ii})] Its negation, denoted by $ \neg b_K $ is given by
$$
\neg b_K \left( {\bf S} \right) = b_K \left( {\bf S} \right) + 1 . \hskip1cm \forall \ {\bf S} \in \mathbb{Z}_2^K \eqno(3)
$$

\end{itemize}

\noindent
{\it N.B.} There is a {\bf total order} among the inputs $ {\bf S} \in \mathbb{Z}_2^K $ of $ b_K $; which we are going to follow along this work. It is given by the bijection
$ \mathbb{Z}_2^K \stackrel{s}{\longleftrightarrow} \left[ 2^K \right] $, given by
$$
s \left( {\bf S} \right) = 1 + \sum_{i=1}^K \ S_i \ 2^{i-1} \hskip1.0cm 1
\leq s \left( {\bf S} \right) \leq 2^K . \eqno(4)
$$

\begin{itemize}

\item[] {\bf Definition}~2:

A $K$-Boolean function (1) is completely determined by its {\it truth table}
$ \mathfrak{B} \left( b_K \right) $, given by
$$
\mathfrak{B} \left( b_K \right) = \left[ \sigma_1, \sigma_2, \dots, \sigma_{2^K}
\right], \eqno(5)
$$
where, $ \sigma_s \in \mathbb{Z}_2 $, is the $s$-th image of (2), under the total order (4).

\end{itemize}

There are $2^{2^K}$ $K$-truth tables $ \mathfrak{B} \left( b_K \right) $, (5)
corresponding to the possible Boolean functions (1). Thus $ \mathfrak{B} $ defines a bijection and we make the

\begin{itemize}

\item[] {\bf Definition}~3:

The set of {\it all $K$-Boolean functions} is given by
$$
\Xi_K = \left\{ b_K : \mathbb{Z}_2^K \longrightarrow \mathbb{Z}_2 \right\} \stackrel{\mathfrak{B}}{\longleftrightarrow} \mathbb{Z}_2^{2^K} .
$$

\end{itemize}

It can also be given a {\bf total order} to the elements of $ \Xi_K $ by the
bijection $ \mathbb{Z}_2^{2^K} \stackrel{\mu}{\longleftrightarrow} \left[ 2^{2^K} \right] $, defined by
$$
\mu \left( b_K \right) = 1 + \sum_{s=1}^{2^K} 2^{s-1} \sigma_s \hskip1.0cm 1
\leq \mu \left( b_K \right) \leq 2^{2^K} , \eqno(6)
$$
which is usually called {\it Wolfram's classification}~${}^{5}$; and is the one used in
Refs.~[9,10,11].

An important function associated to the elements of $ \Xi_K $ is the
{\it weight function} $ \omega: \Xi_K \longrightarrow \left[ 2^K \right] \cup \left\{ 0 \right\} $ defined by
$$
\omega \left( b_K \right) = \sum_{s = 1}^{2^K} \, \sigma_s , \eqno(7)
$$
with their values denoted by $ \omega $. The weight function appears,
mainly, when stochastic extraction of the functions $ b_K $
is involved, as is the case with \textit{NK}-Kauffman networks. There,
each $ \sigma_s \in \mathbb{Z}_2 $ of the truth table (5) is extracted with a {\it bias} probability $ p $ ($ 0 \leq p \leq 1 $) that $ \sigma_s = 1 $, so $ b_K $ has a probability
$$
\Pi \left( b_K \right) \equiv \Pi \left( \omega \right) = p^\omega \, \left( 1 - p \right)^{2^K - \omega}
\eqno(8)
$$
of being extracted~${}^{9,10}$. Due to this important application the weight function is going to be considered further in the work.

We clarify our notation by the example of the truth table (5), for
the case $ K = 2 $; presented in Table~{\it 1}. In the first column, the images
$ \sigma_s $ of the Boolean functions are arranged according to the total
order of their two inputs $ (S_1, S_2) $ given by (4). The next columns
enumerate them according to their Wolfram's number (6), which is
indicated by numbers in boldface. At the bottom of the Table, $ \mathfrak{F} $
stands for the logical meaning of each $2$-Boolean function. For
$ \mu = {\bf 11}, {\bf 13} $, $ \mathfrak{F} = \iota_i = S_i $ ($ i = 1,2 $)
stands for the identity $2$-Boolean function in the $i$-th argument, while
for $ \mu = {\bf 4}, {\bf 6} $, $ \mathfrak{F} = \neg \iota_i = \neg S_i $ represent
their negations (3). The parameter $ \lambda $, of each Boolean function, to
be defined shortly is its {\it degree of irreducibility}; while $ \omega $,
its weight; given by (7).

\footnotesize
\begin{center}
\begin{tabular}{|c|c|c|c|c|c|c|c|c|c|c|c|c|c|c|c|c|}
\hline $ \mathfrak{B}(b_2) $ & {\bf 1} & {\bf 2}
& {\bf 3} & {\bf 4} & {\bf 5} & {\bf 6} & {\bf 7} & {\bf 8} & {\bf
9} & {\bf 10} & {\bf
11} & {\bf 12} & {\bf 13} & {\bf 14} & {\bf 15} & {\bf 16} \\
\hline $ \sigma_1  $ & 0 & 1 & 0 & 1 & 0 & 1 & 0 & 1 & 0 & 1 & 0 & 1 & 0 & 1 & 0 & 1 \\
\hline $ \sigma_2  $ & 0 & 0 & 1 & 1 & 0 & 0 & 1 & 1 & 0 & 0 & 1 & 1 & 0 & 0 & 1 & 1 \\
\hline $ \sigma_3  $ & 0 & 0 & 0 & 0 & 1 & 1 & 1 & 1 & 0 & 0 & 0 & 0 & 1 & 1 & 1 & 1 \\
\hline $ \sigma_4  $ & 0 & 0 & 0 & 0 & 0 & 0 & 0 & 0 & 1 & 1 & 1 & 1 & 1 & 1 & 1 & 1 \\
\hline $-$ & $-$ & $-$ & $-$ & $-$ & $-$ & $-$ & $-$ & $-$ & $-$ & $-$ & $-$ & $-$ & $-$
& $-$ & $-$ & $-$ \\
\hline $ {\mathfrak F} $ & $ \neg \tau $  & $\neg \vee$ & $ \nRightarrow $  &
$ \neg \iota_2 $  & $ \nLeftarrow $ & $ \neg \iota_1 $ & $ \nLeftrightarrow $ & $ \neg
\wedge $ & $ \wedge $  & $ \Leftrightarrow $ & $ \iota_1 $ & $ \Leftarrow $ & $ \iota_2 $
& $ \Rightarrow $ & $ \vee $ & $ \tau $  \\
\hline $ \lambda  $ & 0 & 2 & 2 & 1 & 2 & 1 & 2 & 2 & 2 & 2 & 1 & 2 & 1 & 2
& 2 & 0 \\ \hline $ \omega  $ & 0 & 1 & 1 & 2 & 1 & 2 & 2 & 3 & 1 & 2 & 2 & 3
& 2 & 3 & 3 & 4 \\
\hline

\end{tabular}
\end{center}
\normalsize \baselineskip = 14.8pt

\centerline{\textbf{Table~1.} The $ \mathfrak{B}(b_2) $ truth tables of the
sixteen $2$-Boolean functions.}

\

Not all the $K$-Boolean functions depend completely on their $K$ arguments.
Extreme examples, for $ K = 2 $, from Table~{\it 1}: are rules {\bf 1} and {\bf 16}
({\it contradiction} and {\it tautology}, respectively), which do not depend
on either $ S_1 $ or $ S_2 $. Rules {\bf 4}, {\bf 6}, {\bf 11} and {\bf 13}
only depend on one of the arguments; while the remaining $10$ depend on both.
It has been shown that these facts are responsible of important transition behaviors in \textit{NK}-Kauffman networks, with repercussions in their applications
to biology~${}^{9-11}$. Then, a classification in terms of what is going to be called the {\it irreducible-degree} of $ b_k $; was proposed in Ref.~[9].

Let us make the following

\


\leftline{{\bf Definitions 4:}}

\begin{itemize}

\item[{\it i})] $ \forall $ $ i \in [K] $,  $ b_K $ is $\{i\}$-{\it irreducible}
$ \Leftrightarrow \exists $ $ {\bf S} = \left( S_1, \dots, S_K \right) \in \mathbb{Z}_2^K $:
$$
b_K \left( S_1, \dots, S_i, \dots, S_K \right) = 1 + b_K \left( S_1,
\dots, S_i + 1, \dots, S_K \right).
$$
On the contrary, if $ \forall $ $ {\bf S} \in \mathbb{Z}_2^K $
$$
b_K \left( S_1, \dots, S_i, \dots, S_K \right) = b_K \left( S_1,
\dots, S_i + 1, \dots, S_K \right);
$$
$ b_K $ is $\{i\}$-{\it reducible}.

\item[{\it ii})] $ \forall $ $ \left\{ i_1, \dots, i_\lambda \right\}
\subseteq [K] $, $ b_K $ is $\left\{ i_1, \dots, i_\lambda \right\}$-{\it irreducible}
 if it is $\{i_\alpha\}$-{\it irreducible} for $ \alpha = 1, \dots, \lambda $.
Similarly, $ b_K $ is $\left\{ i_1, \dots, i_\lambda \right\}$-{\it reducible}.

\item[{\it iii})] $ b_K $ is {\it irreducible of degree} $ \lambda $
($ \lambda = 0, 1, \dots, K $); if it is irreducible over $ \lambda
$ indexes $ \left\{ i_1, \dots, i_\lambda \right\} \subseteq \left[ K \right] $ and reducible on the remaining $ K - \lambda $ indexes $ [K] \setminus \left\{ i_1, \dots, i_\lambda \right\} $.

\item[{\it iv})] For $ \lambda = K $, $ b_K $ is called to be
{\it totally-irreducible}.

\item[{\it v})] $ \forall \ i \in \left[ K \right] $
$$
\mathfrak{R}_K \left\{i \right\} = \left\{ b_K \in \Xi_K \ | \ b_K \ {\rm is} \
\left\{ i \right\}-{\rm reducible} \right\}
$$

\item[{\it vi})] $ \forall \ \left\{ i_1, \dots, i_\lambda \right\} \subseteq
\left[ K \right] $

$$
\mathfrak{R}_K \left\{ i_1, \dots, i_\lambda \right\} \equiv \bigcap_{\alpha = 1}^\lambda
\mathfrak{R}_K \left\{i_\alpha \right\} \eqno(9)
$$
{\it N.B.} $ \mathfrak{R}_K \left\{ i_1, \dots, i_\lambda \right\} $ carries no
information about whether, their elements, are (or not) reducible on the remaining
$ \left[ K \right] \setminus \left\{ i_1, \dots, i_\lambda \right\} $ indexes.

\item[{\it vii})] $\forall \ b_K \in \Xi_K $, $ \lambda \left( b_K \right) $
represents the function
$$
\lambda: \Xi_K \to \left[K \right] \cup \left\{ 0 \right\} \eqno(10)
$$
that gives the {\it irreducible-degree} of $ b_K $, with $ \lambda $ standing
for their values.

\end{itemize}

From these {\it Definitions} we have the following

\


\leftline{{\bf Decompositions:}}

\begin{itemize}

\item[{\it i})] $ \Xi_K $ is disjointed decomposed in terms of the irreducible-degree $ \left( b_K \right) $ of their elements by
$$
\Xi_K = \bigsqcup_{\lambda = 0}^K \ {\mathfrak T}_K \left( \lambda
\right) , \eqno(11a)
$$
where
$$
{\mathfrak T}_K \left( \lambda \right) = \left\{ b_K \in \Xi_K | \lambda
\left( b_K \right) = \lambda \right\} . \eqno(11b)
$$

\item[{\it ii})] From (9), $ {\mathfrak T}_K \left( \lambda \right) $ may be decomposed in its turn, as a disjoint union over all the indexes $ \left\{ i_1, \dots, i_\lambda \right\} \subseteq \left[ K \right] $ with cardinality $ \lambda $, in the following way:

\item[{\it a})] The $K$-Boolean functions $\left\{ i_1, \dots, i_\lambda \right\}$-{\it irreducible} are given, from {\it Definitions~4 (ii)} and {\it (vi)}, by
    $$
\bigcap_{\alpha = 1}^\lambda \left( \Xi_K \setminus \mathfrak{R}_K \left\{ i_\alpha \right\} \right)  = \Xi_K \setminus
\left( \bigcup_{\alpha = 1}^\lambda \mathfrak{R}_K \left\{ i_\alpha \right\} \right)
$$
where Morgan's Law was used.

\item[{\it b})] The $K$-Boolean functions which are reducible in $ j \in \left[ K \right] \setminus \left\{i_1, \dots, i_\lambda \right\} $ are given by $ \bigcap_j \ \mathfrak{R}_K \left\{ j \right\} $.

Then
$$
{\mathfrak T}_K \left( \lambda \right) = \bigsqcup_{\left\{ i_1, \dots, i_\lambda \right\} \subseteq \left[ K \right]}  \left\{ \ \left[ \Xi_K \setminus
\left( \bigcup_{\alpha = 1}^\lambda \mathfrak{R}_K \left\{ i_\alpha \right\} \right)
\right] \cap \right.
$$
$$
\cap \left. \left[ \bigcap_{j \in \left[ K \right] \setminus \left\{i_1, \dots, i_\lambda \right\}} \mathfrak{R}_K \left\{ j \right\} \right] \ \right\} , \eqno(12)
$$

\item[{\it iii})] $ \Xi_K $ is disjointed decomposed in terms of the weight $ \omega \left( b_K \right) $ of their elements by
$$
\Xi_K = \bigsqcup_{\omega = 0}^{2^K} \ \mathfrak{P}_K \left( \omega
\right), \eqno(13a)
$$
where
$$
\mathfrak{P}_K \left( \omega \right) = \left\{ b_K \in \Xi_K \, | \, \omega \left( b_K \right) = \omega \right\} , \eqno(13b)
$$
with cardinalities
$$
\# \mathfrak{P}_K \left( \omega \right) = {2^K \choose \omega} . \eqno(14)
$$

\end{itemize}

\

A recursive formula for the cardinalities $ \beta_K \left( \lambda \right) \equiv \# {\mathfrak T}_K \left( \lambda \right) $ was found in Ref.~[9] obtaining
$$
\beta_K \left( \lambda \right) = {K \choose \lambda} \
\mathfrak{G}_\lambda , \eqno(15a)
$$
where $ \mathfrak{G}_\lambda \equiv \beta_\lambda \left( \lambda \right) $. Taking cardinalities in decomposition (11), it follows,
$$
2^{2^K} = \sum_{\lambda = 0}^K {K \choose \lambda} \
\mathfrak{G}_\lambda ;
$$
which can be inverted to obtain~${}^{13}$:
$$
\mathfrak{G}_\lambda = \sum_{m = 0}^\lambda \left( -1 \right)^{\lambda - m} \, {\lambda \choose m} \ 2^{2^m} . \eqno(15b)
$$
All the coefficients $ \beta_K \left( \lambda \right) $, but $ \beta_K \left( 0 \right) = 2 $, grow with $K$. Note that $ {\mathfrak T}_K \left( 0 \right) = \left\{ \neg \tau, \tau \right\} $ consists only in the {\it contradiction}, and {\it tautology} functions; and their truth tables (5) are given by
$$
\mathfrak{B}_K \left( \neg \tau \right) = \underbrace{\left[ 0, 0, \dots, 0 \right] }_{2^K} ,
$$
and
$$
\mathfrak{B}_K \left( \tau \right) = \underbrace{\left[ 1, 1, \dots, 1 \right] }_{2^K}.
$$
On the other hand, from (15b) an asymptotic expression, for the number of totally-irreducible functions $ \beta_K(K) = \mathfrak{G}_K $, for $ K \gg 1 $, is obtained with respect to the total number of $K$-Boolean functions $ 2^{2^K} $, giving
$$
{\mathfrak{G}_K \over 2^{2^K}} \approx 1 - {\cal O} \left( {K \over 2^{2^{K-1}}}\right) . \eqno(16)
$$
So, with respect to the normalized counting measure, almost any $K$-Boolean function is totally-irreducible.

Irreducibility in Boolean functions shows us, that the {\bf real connectivity} of a $K$-Boolean function $ b_K $ is not $ K $; but $ \lambda \left( b_K \right) $. So $ \beta_K \left( \lambda \right) $, defined by (15), gives a real gauge for it. When stochastic extraction of  $ b_K $ through (8) is involved, the function $ \omega \left( b_K \right) $ given by (7) norms the average amount of Boolean functions involved in such processes. So, an important quantity to be considered when both reducibility, and stochastic extraction of $ b_K $ must be taken into account; is the joint probability distribution in terms of the irreducible-degree, and the weight, given by $ \varrho_K \left( \lambda, \omega \right) / 2^{2^K} $, where
$$
\varrho_K \left( \lambda, \omega \right) = \# \left[ {\mathfrak T}_K
\left( \lambda \right) \cap \mathfrak{P}_K \left( \omega \right) \right] .
$$
Calculation of $ \varrho_K \left( \lambda, \omega \right) $ is quite involved if
strict use of {\it Decompositions}~(11) and (13), with their cardinalities (14) and (15) are only used. In the following sections, we will construct a ring structure support to deal easily with quantities like this one.


\section{3. $K$-Boolean Functions and their representation in $ {\cal P}^2
\left[ K \right] $.}

It is possible to obtain a better understanding of Boolean irreducibility by recasting the description from {\it Definitions~4}, into set and ring theoretical languages; which will increase considerably our calculation combinatorial counting power of important quantities. More concrete, $ \left( \Xi_K, \, +, \, \cdot \, \right) $ constitutes a Boolean ring, and we are going to find subrings related to sets of reducible functions, to be defined below. Let us see this in detail:

In general, for any set $ \Omega $, its power set  $ \mathcal{P}\,
\Omega  = \left\{ \mathcal{D} \, | \, \mathcal{D} \subseteq
\Omega \right\} $ is a Boolean ring with the set operations {\it symmetrical difference} $ \bigtriangleup $ (addition), {\it intersection} $\cap $ (product), and; with $ \emptyset $ and $ \Omega $ constituting the identical elements under addition and product, respectively~${}^{14}$. There is a ring-isomorphism $ \Phi $ into the set of Boolean functions
$$
2^\Omega \equiv\left\{ \mathfrak{X}: \Omega \rightarrow \mathbb{Z}_2 \right\}
\stackrel{\Phi}{\longrightarrow} \mathcal{P}\, \Omega , \eqno(17)
$$
by the assignment of a {\it characteristic set} $ \mathcal{D}_\mathfrak{X} $ of the boolean function $ \mathfrak{X} $; which is defined by
$$
\mathcal{D}_\mathfrak{X} = \Phi \left(  \mathfrak{X} \right) = \left\{ d \in \Omega \, | \, \mathfrak{X} \left( d \right) = 1 \right\} \subseteq \Omega ;
$$
and has as a unique inverse association the {\it characteristic function} $ \mathfrak{X}_\mathcal{D} $ of the set $ \mathcal{D} $, defined by
$$
\mathfrak{X}_\mathcal{D} \left( d \right) = \left\{ \begin{array}{ll}
1 & \mbox{if $ d \in \mathcal{D} $}
\\ {} & {} \\ 0 & \mbox{if $ d  \notin \mathcal{D} $} \end{array} \right. \hskip1cm \forall \ \mathcal{D} \in \mathcal{P}\, \Omega \ .
$$
So, $ \forall \ \mathcal{D}, \mathcal{F} \in \mathcal{P}\, \Omega $, the
following ring-isomorphic properties are satisfied

\begin{itemize}

\item[] {\it i)} $\mathfrak{X} _{\mathcal{D} \bigtriangleup \mathcal{F}} =
\mathfrak{X}_{\mathcal{D}} + \mathfrak{X} _{\mathcal{F}}$

\item[] {\it ii)} $ \mathfrak{X}_{\mathcal{D} \cap \mathcal{F}} =
\mathfrak{X}_{\mathcal{D}} \, \cdot \, \mathfrak{X}_{\mathcal{F}} $

\item[] {\it iii)} $\mathfrak{X} _{\emptyset} \equiv 0 $,
$\mathfrak{X} _{\Omega} \equiv 1 $.

\end{itemize}

See details in Ref.~[14].

Using the ring-isomorphism (17), for the inputs $ {\bf S} \in \mathbb{Z}_2^K $ of (2), we have
$$
\mathbb{Z}_2^K \cong 2^{\left[ K \right]} = \left\{ \mathfrak{X}: \left[ K \right] \rightarrow \mathbb{Z}_2 \right\} \stackrel{\Gamma}{\longrightarrow}
{\cal P} \left[ K \right]
$$
with the characteristic set $ \mathcal{A}_\mathfrak{X} $ of $ \mathfrak{X} $ given by
$$
\mathcal{A}_\mathfrak{X} = \Gamma \left(  \mathfrak{X} \right) = \left\{ i \in \left[ K \right]
\, | \, \mathfrak{X} \left( i \right) = S_i = 1 \right\} \subseteq \left[ K \right]
, \eqno(18)
$$
where $ S_i $ is given by (2). The bijection (4) establishes a {\it total order} $ s $
\break ($ s = 1, \dots, 2^K $) among the elements of $ {\cal P} \left[ K \right] $. So, we may label them by
$$
\mathcal{A}_s = \left\{ i_1, \dots, i_l \right\} \subseteq \left[ K \right] ,
$$
where: $ i_\alpha \in \left[ K \right] $, $ \alpha \in \left[ l \right] $, and $ l \in \left[ K \right] $.
Tautologically,
$$
s = s \left( \mathcal{A} \right) = 1 + \sum_{i_\alpha \in \mathcal{A}} 2^{i_\alpha - 1} = 1 + \sum_{i=1}^K \, 2^{i-1} \, \mathfrak{X}_\mathcal{A} \left( i \right) . \eqno(19)
$$

Same consideration may now be applied to $ \Xi_K $, with the total order (19); by using once again the ring-isomorphism (17). We obtain;
$$
\Xi_K \cong 2^{{\cal P} \left[ K \right]} = \left\{ \mathfrak{X}: {\cal P} \left[ K \right] \rightarrow
\mathbb{Z}_2 \right\} \stackrel{\Psi}{\longrightarrow} {\cal P}^2
\left[ K \right] \equiv  {\cal P} {\cal P} \left[ K \right] , \eqno(20)
$$
with the characteristic set $ \mathcal{B}_\mathfrak{X} $ of $ \mathfrak{X} $ given by
$$
\mathcal{B}_\mathfrak{X} = \Psi\left(  \mathfrak{X} \right) = \left\{\mathcal{A}_s \in {\cal P} \left[ K \right] \ | \ \mathfrak{X} \left( \mathcal{A}_s \right) =
\sigma_s = 1 \right\} \subseteq \mathcal{P} \left[ K \right] , \eqno(21)
$$
with $ \sigma_s $ given by the truth table (5) of the $K$-Boolean function $ \mathfrak{X} $. Now, using bijection (6), we may use the label
$ \mu $ ($ \mu = 1, \dots, 2^{2^K} $) for the elements of $ {\cal P}^2 \left[ K \right] $, to obtain:
$$
\mathcal{B}_\mu = \left\{ \mathcal{A}_{s_1}, \dots, \mathcal{A}_{s_m} \right\}
\subseteq \mathcal{P} \left[ K \right] ,
$$
where: $ s_\beta \in \left[ 2^K \right] $, $ \beta \in \left[ m \right] $, and $ m \in \left[ 2^K \right] $. Tautologically
$$
\mu = \mu \left( \mathcal{B}\right) \, = \ 1 \ + \sum_{s \in \left\{ s \, | \, \mathcal{A}_s \in \, \mathcal{B} \right\}} 2^{s - 1} \, = \, 1 \, + \, \sum_{s=1}^{2^K} \, 2^{s -1}
\, \mathfrak{X}_{\mathcal{B}} \left( \mathcal{A}_s \right) .
$$
The following associations are going to be done in the future:
$$ \mathcal{B} = \Psi \left( b_K \right) \in \mathcal{P}^2 \left[ K \right] ,
$$
and through (18) and (19)
$$
\mathcal{A}= \Gamma \left( {\bf S} \right) \in \mathcal{P} \left[ K \right]. \eqno(22)
$$
To be noted also from (7) and (21) that
$$
\omega \left( b_K \right) = \# \Psi \left( b_K \right) = \# \mathcal{B} . \eqno(23)
$$

\

\section{4. $ \mathfrak{R}_K \left\{ i_1, \dots, i_\lambda \right\} $ is a Subring of $ \left( \mathcal{P}^2 \left[ K \right], \ \triangle, \ \cap \right) $.}

For $ \mathfrak{R}_K \left\{ i_1, \dots, i_\lambda \right\} $, given by (9), we are going to show that there is a ring-isomorphism $ \mathfrak{R}_K \left\{ i_1, \dots, i_\lambda \right\} \cong \mathcal{P}^2 \left( \left[ K \right] \setminus \left\{ i_1, \dots, i_\lambda \right\} \right)  $; that allows us to easy count weight functions by means of (23). To do so, first we will recast {\it Definitions~4} for the elements of $ \Xi_K $ in the language of the elements of $ \mathcal{P}^2 \left[ K \right] $. Let us begin with the following:

\newpage

\leftline{{\bf Lemmas}}

\begin{itemize}

\item[{\it i.a})] $ \forall \ \mathcal{A} \in \mathcal{P} \left[ K \right] $, and $ \forall \ i \in \left[ K \right] $ $ \Rightarrow $ $ \mathcal{A} \not= \mathcal{A} \, \triangle \left\{ i \right\} $, and $ \left( \, \mathcal{A} \, \triangle \left\{ i \right\} \, \right) \triangle \left\{ i \right\} = \mathcal{A} $.

\item[] {\bf Proof}: Follows from the fact that $ \left\{ i \right\} \neq \emptyset $ and the nilpotent property of the symmetrical difference $ \triangle $.

\rightline{$ \blacksquare $}

\item[{\it i.b})] $ \forall \ \mathcal{A} \in \mathcal{P} \left[ K \right] $, and $ \forall \ i, j \in \left[ K \right] $, such that $ i \not= j \ $ $ \ \Rightarrow $

    $ \mathcal{A} \, \triangle \left\{ i \right\} \, \triangle \left\{ j \right\} = \mathcal{A} \, \triangle \left\{ i, j \right\} $.

\item[] {\bf Proof}: Follows from the fact that $ \left\{ i \right\} \cap \left\{ j \right\} = \emptyset $.

\rightline{$ \blacksquare $}

\item[{\it ii})] Let $ b_K: \mathbb{Z}_2^K \rightarrow \mathbb{Z}_2 $, and $ \mathcal{B} = \Psi \left( b_K \right) \in \mathcal{P}^2 \left[ K \right] $ be its associated set. Then: $ b_K $ is $\left\{ i \right\}$-reducible $ \Longleftrightarrow $ $ \forall \ \mathcal{A} \in \mathcal{P} \left[ K \right] $, $ \ \mathfrak{X}_\mathcal{B} \left( \mathcal{A} \right) = \mathfrak{X}_\mathcal{B} \left( \mathcal{A} \, \triangle \left\{ i \right\} \right) \ $.

\item[] {\bf Proof}: From (22), and {\it Lema~1.a};
$$
\mathcal{A} \, \triangle \left\{ i \right\} = \Gamma \left( S_1, \dots, S_i, \dots, S_K \right) \ \triangle \ \Gamma ( \underbrace{0, \dots, 0, 1}_i, 0, \dots, 0 )
$$
$$
= \Gamma \left( S_1, \dots, S_i + 1, \dots, S_K \right).
$$
Now, from {\it Definition~4.i} the {\it Lemma} follows.

\rightline{$ \blacksquare $}

\end{itemize}

\newpage

\leftline{\bf Corollary I}

\begin{itemize}

\item[] $ \forall \ b_K \in \Xi_K \ \Rightarrow \ \lambda \left( b_K \right) = \lambda \left( \neg b_K \right) $.

\item[] {\bf Proof}: Let $ \mathcal{\tilde{B}} = \Psi \left( \neg b_K \right) $, from (21) $ \mathfrak{X}_\mathcal{\tilde{B}} \left( \mathcal{A} \right) = \mathfrak{X}_\mathcal{B} \left( \mathcal{A} \right) + 1 $, $ \forall \ \mathcal{A} \in \mathcal{P} \left[ K \right] $. So, from {\it Lemma (ii)}; $ b_K $ is $\{i\}$-reducible $ \Longleftrightarrow $ $ \neg b_K $ is $\{i\}$-reducible.

\rightline{$ \blacksquare $}

\end{itemize}

\leftline{{\bf Lemma}}

\begin{itemize}

\item[{\it iii})] With the total order of $ \mathcal{P} \left[ K \right] $ given by (4), or equivalently (19), \break $ \forall \ \mathcal{A} \in \mathcal{P} \left[ K \right] $
$$
s \left( \mathcal{A} \right) = s \left( \mathcal{A} \, \triangle \left\{ i \right\} \right) + 2^{i-1} \, \left( 2 \ \mathfrak{X}_\mathcal{A} \left( i \right) - 1 \right) . \eqno(24)
$$

\item[] {\bf Proof}: $ \forall \ \mathcal{A} \in \mathcal{P} \left[ K \right] $, the inverse image of (22) gives
    $$
    \Gamma^{-1} \left( \mathcal{A}\right) = \left( S_1, \dots, S_i, \dots, S_K \right) , $$
    and
    $$
    \Gamma^{-1} \left( \mathcal{A} \, \triangle \left\{ i \right\}\right) = \left( S_1, \dots, S_i + 1, \dots, S_K \right) .
    $$
    Then from (19)
    $$
    s \left( \mathcal{A} \, \triangle \left\{ i \right\} \right) = 1 + \sum_{\stackrel{j=1}{j \not= i}}^K \ S_j \ 2^{j-1} + 2^{i - 1} \, \left[ S_i + 1 \right]_2
    $$
    $$
    \hskip3.4cm = 1 + \sum_{j=1}^K \ S_j \ 2^{j-1} - 2^{i - 1} \left( \, S_i - \, \left[ S_i + 1 \right]_2 \right)
    $$
    $$
    \hskip1.1cm= s \left( \mathcal{A} \right) - 2^{i-1} \, \left( 2 \ \mathfrak{X}_\mathcal{A} \left( i \right) - 1 \right) .
    $$

\rightline{$ \blacksquare $}

\item[{\it iv})] $ \forall \, \mathcal{B} = \Psi \left( b_K \right) \in \mathcal{P}^2 \left[ K \right] $, with the total order (19), and with the truth table $ \mathfrak{B} \left( b_K \right) $ given by (5):
    $$
    b_K  \ {\rm is} \ \left\{ i \right\}-{\rm reducible}  \ \Longleftrightarrow  \ \forall \, \mathcal{A} \in \mathcal{P} \left[ K \right] \ \sigma_{s \left( \mathcal{A} \right)} = \sigma_{s \left( \mathcal{A} \, \triangle \left\{ i \right\} \right)} \ . \eqno(25)
    $$

\item[] {\bf Proof}: From {\it Lemma (ii)}, and the ring-isomorphic association (20): $ \sigma_{s \left( \mathcal{A} \right)} = \mathfrak{X}_\mathcal{B} \left( \mathcal{A} \right) $, and $ \sigma_{s \left( \mathcal{A} \, \triangle \left\{ i \right\} \right)} = \mathfrak{X}_\mathcal{B} \left( \mathcal{A} \, \triangle \left\{ i \right\} \right) $.

\rightline{$ \blacksquare $}

\end{itemize}

\begin{itemize}

\item[] {\bf Theorem I}: The operational form of the irreducible-degree (10) of $ b_K \in \Xi_K $ is given by
$$
\lambda \left( b_K \right) = \sum_{i=1}^K \ \Theta \left( \, \mathcal{F}_K \left( b_K; i \right) \, \right) , \eqno(26a)
$$
where, for any $ a \in \mathbb{R} $, the step function $ \Theta $ is given by
$$
\Theta \left( a \right) = \left\{ \begin{array}{ll}
1 & \mbox{if $ a > 0 $}
\\ {} & {} \\ 0 & \mbox{if $ a \leq 0 $} \end{array} \right. ,  \eqno(26b)
$$
and
$$
\mathcal{F}_K \left(  b_K; i \right) \equiv \sum_{b=1}^{2^{K-i}} \ \ \sum_{s= \left(b-1\right) 2^i + 1}^{\left(b-1\right) 2^i + 2^{i-1}} \ \ \left[ \sigma_s + \sigma_{\left( s + 2^{i-1}\right)} \right]_2 . \eqno(26c)
$$

\item[] {\bf Proof}: From (24), and (25) follows that if $ b_K $ is $\{ i \} $-{\it irreducible}, there exists an $ \mathcal{A} \in \mathcal{P} \left[ K \right] $ such that $ \sigma_{s \left( \mathcal{A} \right)} \not= \sigma_{s \left( \mathcal{A} \, \triangle \left\{ i \right\} \right) } $. Arranging sets, from (24) through the total order (19), so that, $ \mathcal{A} \prec \mathcal{A} \, \triangle \, \left\{ i \right\} $, whenever $ s \left( \mathcal{A} \right) < s \left( \mathcal{A} \, \triangle \left\{ i \right\} \right) $ it follows that $ s \left( \mathcal{A} \right) = s \left( \mathcal{A} \, \triangle \left\{ i \right\} \right) -2^{i-1} $: So, the double sum checks over the total number of $2^{K-1}$ pairs of indexes $ \left[ s \left( \mathcal{A} \right), s \left( \mathcal{A} \, \triangle \left\{ i \right\} \right) \right] $, if it happens that $ \sigma_{s \left( \mathcal{A} \right)} \not= \sigma_{s \left( \mathcal{A} \, \triangle \left\{ i \right\} \right)} $. If that is the case at least for one of such pairs; then $ \mathcal{F}_K \left(  b_K; i \right) > 0 $, and so $ \Theta \left( \, \mathcal{F}_K \left(  b_K; i \right) \, \right) = 1 $; implying that $ b_K $ is $\{ i \} $-{\it irreducible}.

\rightline{$ \blacksquare $}

\end{itemize}


\leftline{\bf Lemma}

\begin{itemize}

\item[{\it v})] $ \forall \ i \in \left[K \right] $ there is a ring-isomorphism
$$
\mathcal{P}^2 \left( \left[K \right] \setminus \left\{ i \right\} \right) \stackrel{\varphi_i}{\longrightarrow} \mathfrak{R}_K \left\{ i \right\} \subseteq \mathcal{P}^2 \left[K \right]
$$
given by
$$
\mathcal{B} \stackrel{\varphi_i}{\longmapsto} \mathcal{B} \sqcup \mathcal{B}_i ,
$$
where
$$
\mathcal{B}_i \equiv \left\{ \mathcal{A} \, \triangle \left\{ i \right\} \in \mathcal{P} \left[ K \right] \ | \ \mathcal{A} \in \mathcal{B} \right\} , \eqno(27)
$$
and the unique inverse association
$$
\varphi_i^{-1} \left( \mathcal{B} \right) = \left\{ \mathcal{A} \in \mathcal{B} \, | \, i \notin \mathcal{A} \right\} . \eqno(28)
$$

\item[] {\bf Proof}: From {\it Lemma (ii)}; $ \varphi_i \left( \mathcal{B} \right) \in \mathfrak{R}_K \left\{ i \right\} $ for all $ \mathcal{B} \in \mathcal{P}^2 \left( \left[K \right] \setminus \left\{ i \right\} \right) $. So, due to the unique inverse association (28) $ \varphi_i $ is a bijection. Now, from (27) follows directly that $ \forall \ \mathcal{B}, \mathcal{\tilde{B}} \in \mathcal{P}^2 \left( \left[K \right] \setminus \left\{ i \right\} \right) $ $ \Rightarrow $ $ \left( \mathcal{B} \, \triangle \, \mathcal{\tilde{B}} \right)_i = \mathcal{B}_i \, \triangle \, \mathcal{\tilde{B}}_i $, and $ \left( \mathcal{B} \, \cap \, \mathcal{\tilde{B}} \right)_i = \mathcal{B}_i \, \cap \, \mathcal{\tilde{B}}_i $. Since, by (27) and {\it Lemma (i.a)} $ \mathcal{B} \cap \mathcal{B}_i = \emptyset $, it follows that $ \mathcal{B} \sqcup \mathcal{B}_i = \mathcal{B} \, \triangle \, \mathcal{B}_i $. Then the ring operations $ \triangle $ and $ \cap $ are easily handled to show that $ \varphi_i \left( \mathcal{B} \, \triangle \, \mathcal{\tilde{B}}\right) = \varphi_i \left( \mathcal{B} \right) \, \triangle \, \varphi_i ( \mathcal{\tilde{B}} ) $, and $ \varphi_i \left( \mathcal{B} \, \cap \, \mathcal{\tilde{B}} \right) = \varphi_i \left( \mathcal{B} \right) \, \cap \, \varphi_i ( \mathcal{\tilde{B}} ), \forall \ \mathcal{B}, \mathcal{\tilde{B}} \in \mathcal{P}^2 \left( \left[K \right] \setminus \left\{ i \right\} \right) $.

\rightline{$ \blacksquare $}

\end{itemize}


\leftline{\bf Corollary II}

\begin{itemize}

\item[]
$$ \# \mathfrak{R}_K \left\{ i \right\} = \# \mathcal{P}^2 \left( \left[K \right] \setminus \left\{ i \right\} \right) = 2^{2^{K-1}} .
$$

\rightline{$ \blacksquare $}

\item[] {\bf Theorem II}: $ \forall \ \left\{ i_1, \dots, i_\lambda \right\} \in \mathcal{P}\left[K \right] $ there is a ring-isomorphism
$$
\varphi_{i_1, \dots, i_\lambda}: \mathcal{P}^2 \left( \left[K \right] \setminus \left\{ i_1, \dots, i_\lambda \right\} \right) \longrightarrow \mathfrak{R}_K \left\{ i_1, \dots, i_\lambda \right\} \subseteq \mathcal{P}^2 \left[K \right]
$$
given by
$$
\varphi_{i_1, \dots, i_\lambda} \left( \mathcal{B} \right) =  \mathcal{B} \sqcup \left( \bigsqcup_{\left\{ i_\alpha \right\}} \mathcal{B}_{ i_\alpha} \right) \sqcup \left( \bigsqcup_{\left\{ i_{\alpha_1}, i_{\alpha_2} \right\}} \mathcal{B}_{ i_\alpha, i_\beta} \right) \sqcup \dots
$$
$$
\sqcup \left( \bigsqcup_{\left\{ i_{\alpha_1}, \dots, i_{\alpha_j} \right\}}  \mathcal{B}_{ i_{\alpha_1}, \dots, i_{\alpha_j}} \right) \sqcup \dots \sqcup \mathcal{B}_{ i_1, \dots, i_\lambda} , \eqno(29)
$$
where $ \left\{ i_{\alpha_1}, \dots, i_{\alpha_j} \right\} \subseteq \left\{ i_1, \dots, i_\lambda \right\} $ ($ \alpha_1 < \dots < \alpha_j $) runs over all the subsets of  $ \left\{ i_1, \dots, i_\lambda \right\} $ in cardinal order $ j = 1, \dots, \lambda $, and,
$$
\mathcal{B}_{i_{\alpha_1}, \dots, i_{\alpha_j}} \equiv \left\{ \mathcal{A} \, \triangle \left\{i_{\alpha_1}, \dots, i_{\alpha_j} \right\} \in \mathcal{P} \left[ K \right] \ \ | \ \mathcal{A} \in \mathcal{B} \right\} . \eqno(30)
$$
With the inverse given by
$$
\varphi_{i_1, \dots, i_\lambda}^{-1} \left( \mathcal{B} \right) = \left\{ \mathcal{A} \in \mathcal{B} \, | \, i_\alpha \notin \mathcal{A} \ \ \forall \, \alpha = 1, \dots, \lambda \right\} .
$$

\item[] {\bf Proof}: From (30) and {\it Lemmas (i)}, $ \forall \ \mathcal{B} \in \mathcal{P}^2 \left( \left[ K \right] \setminus \left\{ i_1, \dots, i_\lambda \right\} \right) $

$$
\mathcal{B} \cap \mathcal{B}_{i_{\alpha_1}, \dots, i_{\alpha_j}} = \mathcal{B}_{i_{\alpha_1}, \dots, i_{\alpha_j}} \cap \mathcal{B}_{i_{\beta_1}, \dots, i_{\beta_k}} = \emptyset ,
$$
$ \forall \ \left\{ i_{\alpha_1}, \dots, i_{\alpha_j} \right\} \neq \left\{ i_{\beta_1}, \dots, i_{\beta_k} \right\} $. So the union is disjoint. By induction from {\it Lemma (v)}, aided by {\it Lemma (i.b)}, and the property that

$ \mathcal{B} \in \mathfrak{R}_K \left\{ i_1, \dots, i_\lambda \right\} \ \Longleftrightarrow $
$$
\left[ \ \forall \ \mathcal{A} \in \mathcal{B} \Longleftrightarrow \ \mathcal{A} \, \triangle \ \mathcal{C} \in \mathcal{B}; \ \ \forall \ \mathcal{C} \subseteq \left\{ i_1, \dots, i_\lambda \right\} \ \right] ;
$$
which follows from {\it Lemmas (i.b)} and {\it (ii)}; the {\it Theorem} is proved.

\rightline{$ \blacksquare $}

\end{itemize}

\leftline{\bf Corollary III}

\begin{itemize}

\item[]
$$ \# \mathfrak{R}_K \left\{ i_1, \dots, i_\lambda \right\} = \# \mathcal{P}^2 \left( \left[K \right] \setminus \left\{ i_1, \dots, i_\lambda \right\} \right) = 2^{2^{K-\lambda}} .
$$

\rightline{$ \blacksquare $}

\end{itemize}

\newpage

\leftline{\bf Lemma}

\begin{itemize}

\item[{\it vi})] There is a bijection

\end{itemize}

$$
\mathfrak{R}_K \left\{ i_1, \dots, i_\lambda \right\} \cap \mathfrak{P}_K \left( \omega \right) \cong \left\{  \mathcal{B} \in \mathcal{P}^2 \left( \left[ K \right] \setminus \left\{ i_1, \dots, i_\lambda \right\} \right) \ | \ \# \mathcal{B} = {\omega \over 2^\lambda} \right\} \ .
$$

\begin{itemize}

\item[] {\bf Proof}: From (30); $ \# \mathcal{B} = \# \mathcal{B}_{i_{\alpha_1}, \dots, i_{\alpha_j}} $ $ \forall  \ \left\{ i_{\alpha_1}, \dots, i_{\alpha_j} \right\} \subseteq \left\{ i_1, \dots, i_\lambda \right\} $. Taking cardinalities in the disjoint union (29) we have

\end{itemize}

$$
\# \varphi_{i_1, \dots, i_\lambda} \left( \mathcal{B} \right) = \# \mathcal{B} \ \sum_{j=0}^\lambda \ {\lambda \choose j} = \# \mathcal{B} \ 2^\lambda \ .
$$
So $ \varphi_{i_1, \dots, i_\lambda}^{-1} $ contracts cardinalities, and from (23), the weights by a factor $ 2^{- \lambda} $. From (13b) the {\it Lemma} follows.

\rightline{$ \blacksquare $}


\leftline{\bf Corollary IV}

\begin{itemize}

\item[]

$$
\# \ \left[ \ \mathfrak{R}_K \left\{ i_1, \dots, i_\lambda \right\} \cap \mathfrak{P}_K \left( \omega \right) \ \right] = {2^{K-\lambda} \choose \left\lfloor {\omega \over 2^\lambda} \right\rfloor} \ \delta \left( \left\lfloor {\omega \over 2^\lambda} \right\rfloor - {\omega \over 2^\lambda} \right) ,
$$

\end{itemize}

where, $ \forall \ a \in \mathbb{R} $

$$
\delta \left( a \right) = \left\{ \begin{array}{ll}
1 & \mbox{if $ a = 0 $}
\\ {} & {} \\ 0 & \mbox{if $ a \neq 0 $} \end{array} \right.
$$
is Kronecker's delta, and $ \lfloor a \rfloor \in \mathbb{Z} $ the {\it floor function},
defined as the greatest integer $ \lfloor a \rfloor $; such that $ \lfloor a \rfloor \leq a $.

\begin{itemize}

\item[] {\bf Proof}: For any set $ \mathcal{B} $ it happens that $ \# \ \mathcal{B} \in \mathbb{N} \, \cup \, \left\{ 0 \right\} $. From {\it Lemma (vi)}:
    $$
    \mathfrak{R}_K \left\{ i_1, \dots, i_\lambda \right\} \cap \mathfrak{P}_K \left( \omega \right) = \emptyset
    $$
    whenever $ \omega / 2^\lambda \notin \mathbb{N} \, \cup \, \left\{ 0 \right\} $. Otherwise $ \forall \ \mathcal{\tilde{B}} \in \mathfrak{R}_K \left\{ i_1, \dots, i_\lambda \right\} \cap \mathfrak{P}_K \left( \omega \right) $ $ \Rightarrow $ $ \exists \ $ $ \mathcal{B} \subseteq \mathcal{P} \left( \left[ K \right] \setminus \left\{ i_1, \dots, i_\lambda \right\} \right) $, such that $ \# \, \mathcal{B} = \omega / 2^\lambda $ with $ \mathcal{\tilde{B}} = \varphi_{i_1, \dots, i_\lambda} \left( \mathcal{B} \right) $.

\end{itemize}

\rightline{$ \blacksquare $}


\begin{itemize}

\item[] {\bf Theorem III}: $ \varrho_K \left( \lambda, \omega \right) = \# \left[ \ \mathfrak{T}_K \left( \lambda \right) \, \cap \, \mathfrak{P}_K \left( \omega \right) \ \right] $ is analytically given by:
$$
\varrho_K \left( \lambda, \omega \right) = {K \choose \lambda} \ \sum_{m=0}^\lambda \, \left( -1 \right)^{\lambda - m} \ {\lambda \choose m} \times \hskip4cm
$$
$$
\ \times \ {2^m \choose \left\lfloor {\omega \over 2^{K-m}} \right\rfloor} \ \ \delta \left( \left\lfloor {\omega \over 2^{K-m}} \right\rfloor - {\omega \over 2^{K-m}} \right) ,  \eqno(31)
$$

\noindent
where $ 0^0 \equiv 1 $.

\end{itemize}

\begin{itemize}

\item[] {\bf Proof}: Taking the intersection of (12) and (13b), and using the idempotent property of set's intersection ($ \mathcal{D} \cap \mathcal{D} = \mathcal{D} $ for any set $ \mathcal{D} $) we obtain

\end{itemize}


$$
\mathfrak{T}_{K} \left(\lambda \right) \cap \mathfrak{P}_{K} \left( \omega \right) = \bigsqcup_{\left\{ i_1, \dots, i_\lambda \right\} \subseteq \left[ K \right]} \left\{  \ \left[ \Xi_K \setminus
\left( \bigcup_{\alpha = 1}^\lambda \mathfrak{R}_K \left\{ i_\alpha \right\} \right)
\right] \cap   \right.
$$
$$
\cap \ \mathfrak{P}_K \left( \omega\right) \ \cap \left. \left[ \bigcap_{j \in \left[ K \right] \setminus \left\{i_1, \dots, i_\lambda \right\}} \mathfrak{R}_K \left\{ j \right\} \cap \mathfrak{P}_K \left( \omega\right) \right] \ \right\} =
$$
$$
 = \bigsqcup_{\left\{ i_1, \dots, i_\lambda \right\} \subseteq \left[ K \right]} \left\{ \ \left[ \mathfrak{P}_K \left( \omega\right) \setminus
\left( \bigcup_{\alpha = 1}^\lambda \mathfrak{R}_K \left\{ i_\alpha \right\} \cap \mathfrak{P}_K \left( \omega \right) \right)
\right] \ \cap \right.
$$
$$
\cap \left. \left[ \bigcap_{j \in \left[ K \right] \setminus \left\{i_1, \dots, i_\lambda \right\}} \mathfrak{R}_K \left\{ j \right\} \cap \mathfrak{P}_K \left( \omega\right) \right] \ \right\} =
$$


$$
 = \bigsqcup_{\left\{ i_1, \dots, i_\lambda \right\} \subseteq \left[ K \right]}  \left\{ \ \left[ \bigcap_{j \in \left[ K \right] \setminus \left\{i_1, \dots, i_\lambda \right\}} \mathfrak{R}_K \left\{ j \right\} \cap \mathfrak{P}_K \left( \omega\right) \right] \right.
$$
$$
\setminus \left. \left[ \bigcup_{\alpha = 1}^\lambda \mathfrak{P}_K \left( \omega \right) \cap \mathfrak{R}_K \left\{ i_\alpha \right\} \ \cap \bigcap_{j \in \left[ K \right] \setminus \left\{i_1, \dots, i_\lambda \right\}} \mathfrak{R}_K \left\{ j \right\} \right] \ \right\}
$$

$$
\equiv \bigsqcup_{\left\{ i_1, \dots, i_\lambda \right\} \subseteq \left[ K \right]} \left\{ \left[ {\bf 1} \right]_{i_1, \dots, i_\lambda} \setminus \ \left[ {\bf 2} \right]_{i_1, \dots, i_\lambda} \right\} . \eqno(32a)
$$
From {\it Theorem~II}
$$
\mathcal{M}_{i_1, \dots, i_\lambda} \equiv \bigcap_{j \in \left[ K \right] \setminus \left\{ i_1, \dots, i_\lambda \right\}} \mathfrak{R}_K \left\{ j \right\} \cong \mathcal{P}^2 \left\{ i_1, \dots, i_\lambda \right\} ,
$$
from {\it Lemma (vi)}
$$
\mathfrak{P}_K \left( \omega \right) \cap \mathcal{M}_{i_1, \dots, i_\lambda}\cong \left\{ \mathcal{B} \in \mathcal{P}^2 \left( \left\{ i_1, \dots, i_\lambda \right\} \right) \ | \ \# \mathcal{B} = {\omega \over 2^{K - \lambda}} \right\} ,
$$
and from {\it Corollary IV};
$$
\# \left[ {\bf 1} \right]_{i_1, \dots, i_\lambda} = {2^\lambda \choose \left\lfloor {\omega \over 2^{K-\lambda}} \right\rfloor} \ \ \delta \left( \left\lfloor {\omega \over 2^{K-m}} \right\rfloor - {\omega \over 2^{K-m}} \right) . \eqno(32b)
$$
Since, $ \left[ {\bf 2} \right]_{i_1, \dots, i_\lambda} $ is a union of, non necessarily, disjoint sets; we have~${}^{14}$
$$
\# \left[ {\bf 2} \right]_{i_1, \dots, i_\lambda} = \sum_{n=1}^\lambda \left( -1 \right)^{n-1} \times \hskip9cm
$$
$$
\times \ \sum_{\left\{i_{\alpha_1}, \dots, i_{\alpha_n} \right\} \subseteq \left\{ i_1, \dots, i_\lambda \right\}} \ \# \left[ \ \bigcap_{j=1}^n \mathfrak{R}_K \left\{ i_{\alpha_j} \right\} \cap \ \mathcal{M}_{i_1, \dots, i_\lambda} \cap \mathfrak{P}_K \left( \omega \right) \ \right] . \eqno(32c)
$$
Now
$$
\mathfrak{S}_{i_{\alpha_1}, \dots, i_{\alpha_n}}^{i_1, \dots, i_\lambda} \equiv \bigcap_{j=1}^n \mathfrak{R}_K \left\{ i_{\alpha_j} \right\} \cap \mathcal{M}_{i_1, \dots, i_\lambda} = \bigcap_{j \in \left( \left[ K \right] \setminus \left\{ i_1, \dots, i_\lambda \right\} \right) \cup \left\{ i_{\alpha_1}, \dots, i_{\alpha_n} \right\} } \mathfrak{R}_K \left\{ j \right\}
$$
$$
= \bigcap_{j \in \left[ K \right] \setminus \left( \left\{ i_1, \dots, i_\lambda \right\} \setminus \left\{ i_{\alpha_1}, \dots, i_{\alpha_n} \right\} \right) } \mathfrak{R}_K \left\{ j \right\} \cong \mathcal{P}^2 \left( \left\{ i_1, \dots, i_\lambda \right\} \setminus \left\{ i_{\alpha_1}, \dots, i_{\alpha_n} \right\} \right) ,
$$
where {\it Theorem~II} has been used; and it is worthwhile to bear in mind that $ \left\{ i_{\alpha_1}, \dots, i_{\alpha_n} \right\} \subseteq \left\{ i_1, \dots, i_\lambda \right\} $. From {\it Lemma (vi)} and {\it Corollary IV} we obtain
$$
\# \left[ \mathfrak{S}_{i_{\alpha_1}, \dots, i_{\alpha_n}}^{i_1, \dots, i_\lambda} \cap \mathfrak{P}_K \left( \omega \right) \right] = {2^{\lambda-n} \choose \left\lfloor {\omega \over 2^{K-\lambda+n}} \right\rfloor} \ \ \delta \left( \left\lfloor {\omega \over 2^{K-\lambda+n}} \right\rfloor - {\omega \over 2^{K-\lambda+n}} \right) .
$$
Going to (32a), (32b), (32c), and taking into account that the cardinal counting result is independent of $ \left\{ i_1, \dots, i_\lambda \right\} \subseteq \left[ K \right] $; an overall factor $ {K \choose \lambda} $ is obtained in the union (32a).

\rightline{$ \blacksquare $}

\

\begin{itemize}

\item[] {\bf Checks}: The following formulas, for the number of $ \lambda $-irreducible functions with weight $ \omega $, come as a result of consistency of (31) with (14), and (15):
$$
\sum_{\lambda=0}^K \ \varrho_K \left( \lambda, \omega \right) = {2^K \choose \omega} \ \ {\rm and} \ \ \sum_{\omega=0}^{2^K} \ \varrho_K \left( \lambda, \omega \right) = \beta_K \left( \lambda \right) \ .
$$
See Appendix~B of Ref.~[10] for details on calculations.

\end{itemize}

\begin{itemize}

\item[] {\bf Special Values and Properties}:

\item[{\it 1)}]
$$
\varrho_K \left( 0, \omega \right) = \delta \left( \omega \right) + \delta \left( \omega - 2^K \right) ,
$$
and
$$
\varrho_K \left( \lambda, 0 \right) = \varrho_K \left( \lambda, 2^K \right) = \delta \left( \lambda \right) .
$$
That is; the only completely reducible $K$-Boolean functions are the {\it contradiction} $ \neg \tau $ and {\it tautology} $ \tau $ functions, and also are the only two that have the most extreme values of $ \omega $: $ 0 $ and $ 2^K $ respectively.

\item[{\it 2)}]
$$
\varrho_K \left( 1, \omega \right) = 2 K \ \delta \left( \omega - 2^{K-1} \right) .
$$
Which corresponds to the $ K $-{\it identities} $ \iota_i \equiv b_K \left( S_1, \dots, S_K \right) = S_i $ associated to each one of the arguments $ S_i \in \mathbb{Z}_2 $, $ i = 1, \dots, K $; and the corresponding $K$ negations $ \neg \iota_i = \neg b_K \left( S_1, \dots, S_K \right) = S_i + 1 $.

\item[{\it 3)}] $ \varrho_K \left( \lambda, \omega \right) $ is a symmetrical function of $ \omega $ at the value $ 2^{K-1} $ {\it i.e.}:
$$
\varrho_K \left( \lambda, \omega \right) = \varrho_K \left( \lambda, 2^K - \omega \right) . \eqno(33)
$$

\item[] {\bf Proof}: From the definition (7) of $ \omega ( b_K ) $,  $ \ \omega ( \neg b_K ) = 2^K - \omega ( b_K ) $. From {\it Corollary I}, $ \ \lambda ( \neg b_K ) = \lambda ( b_K ) $, then: $ \forall \ b_K \in \Xi_K $:
$$
b_K \in \mathfrak{T}_K \left( \lambda \right) \cap \mathfrak{P}_K \left( \omega \right) \Longleftrightarrow \ \neg b_K \in \mathfrak{T}_K \left( \lambda \right) \cap \mathfrak{P}_K \left( 2^K - \omega \right)
$$

\rightline{$ \blacksquare $}

\item[{\it 4)}] For $ \omega = 2 n - 1 , n \in \mathbb{N} $,
$$
\varrho_K \left( \lambda, 2 n - 1 \right) = {2^K \choose 2 n - 1 } \ \delta \left( K - \lambda \right) . \eqno(34)
$$

\item[] {\bf Proof}: Follows directly from (31).

\rightline{$ \blacksquare $}

\item[{\it 5)}] For $ \omega = 2 n $, $ n \in \mathbb{N} $, and $ K \gg 1 $,
$$
\varrho_K \left( K, 2 n \right) \approx {2^K \choose 2 n} \ \left[ 1 -  \mathfrak{A} \left(K, n \right) \right]
$$
where the function $ \mathfrak{A} \left(K, n \right) $ goes to zero faster than $ o \left( {K \over 2^K} \right) $ for $ n \sim \mathcal{O} \left( 1 \right) $ as $ K $ grows; and faster than $ o \left( {K \over 2^{2^{K-1}}} \right) $ for, $ n \sim 2^{K-2} $ (in the region of the maximum). So, for $ K \gg 1 $, $ \varrho_K \left( K, 2 n \right) $ obeys a Gaussian probability distribution with the same moments as for the {\bf odd} case (34).

\item[] {\bf Proof}: From (31) with $ \lambda = K $ the leading term is $ {2^K \choose 2 n} $. Of the remaining $ K $ terms the next one in size is $ {2^{K-1} \choose n} $. Using Stirling's approximation for the factorials
$$
{2^{K-1} \choose n} \div {2^K \choose 2 n} \approx \left\{ \begin{array}{ll}
o \left( {1 \over 2^K} \right) & \mbox{for $ n \sim \mathcal{O} \left( 1 \right) $}
\\ {} & {} \\ o \left( {1 \over 2^{2^{K-1}}} \right) & \mbox{for $ n \sim 2^{K-2} $} \end{array} \right.
$$
each one of the remaining $ K -1 $ terms giving smaller contributions.

\rightline{$ \blacksquare $}

\item[] {\bf Corollary V}

\item[] Let
$$
\mathfrak{I}_K \equiv \left\{ b_K \in \Xi_K \, | \, \omega \left( b_K \right) = 2 n - 1, n \in \mathbb{N} \right\} ,
$$
then
$$
\mathfrak{I}_K \varsubsetneq \mathfrak{T}_K \left( K \right) .
$$

\item[] {\bf Proof}: Follows directly from {\it Theorem~III}.

\end{itemize}

\noindent
That is, any $K$-Boolean function $ b_K $ with an odd weight is totally-irreducible. {\it N.B.} The converse is not true, for example, in the case $ K = 2 $ of Table~{\it 1}, the $2$-Boolean functions $ {\bf 7} \, = \, \nLeftrightarrow $, and $ {\bf 10} \, = \, \Leftrightarrow $, both have $ \omega = 2 $ and are totally-irreducible; while the other eight totally-irreducible $2$-Boolean functions have odd weights. This by no means implies that the great majority of totally-irreducible functions have odd weights for growing $K$. Indeed from (34)
$$
\# \mathfrak{I}_K = 2^{2^K - 1} = {1 \over 2} \, \# \, \Xi_K .
$$
So, half of all the $K$-Boolean functions are totally-irreducible with an {\bf odd} weight. Now, for the complement $ \mathfrak{T}_K \left( K \right) \setminus \mathfrak{J}_K $ (the totally-irreducible functions with {\bf even} weights):

Since $ \mathfrak{G}_K $ given by (15b) has the asymptotic behavior (16) for $ K \gg 1 $;
$$
{\# \left( \mathfrak{T}_K \left( K \right) \setminus \mathfrak{J}_K  \right) \over 2^{2^K}} \approx {1 \over 2} - {\cal O} \left( {K \over 2^{2^{K-1}}}\right) .
$$
So, with respect to the normalized counting measure, {\bf almost half} of the totally-irreducible $K$-Boolean function have an {\bf even} weight.

Normalizing according to $ 2^{2^K -1 } $, totally-irreducible $K$-Boolean functions with {\bf odd} weights distribute by a binomial distribution with mean $ \left< \omega \right> = 2^{2^K -1 } $ [ as needed for consistency with (33)], and a standard deviation $ \Sigma^2 = 2^{K-2} $; so the relative error $ \Sigma / \left< \omega \right> = 2^{-K/2} $, decays exponentially to zero with growing $K$. For $ K \gg 1 $ the binomial becomes asymptotically a gaussian distribution, with the very same moments.

\section{5. Conclusions}

In Ref.~[9] a classification of Boolean functions, in terms of their irreducible-degree of connectivity, was constructed allowing to calculate exact, and asymptotic behavior of \textit{NK}-Kauffman networks. In this work we have developed further the mathematical structure and consequences of this classification with the aid of the well known ring-isomorphism (17) {\it i.e.} $ 2^\Omega \, \cong \, \mathcal{P} \, \Omega $~${}^{14}$. In {\it Theorem~I} we found an operational formula (26) for the irreducible-degree $ \lambda \left( b_K \right) $ of a $K$-Boolean function that allows concrete analytical manipulations of it for calculational purposes.

In {\it Theorem~II}, we established a ring-isomorphism between the set \break $ \mathfrak{R}_K \left\{ i_1, \dots, i_\lambda \right\} $ of reducible $K$-Boolean functions on the indexes $ \left\{ i_1, \dots, i_\lambda \right\} \subseteq \left[ K \right] $ and the double power set $ \mathcal{P}^2 \left( \left[K \right] \setminus \left\{ i_1, \dots, i_\lambda \right\} \right) $. This fundamental theorem allows to use the rich ring-isomorphic structure for analytical calculations over Boolean functions. In particular, we have used it to prove {\it Theorem~III} which gives a formula for $ \varrho_K \left( \lambda, \omega \right)$: the number of $K$-Boolean functions with irreducible-degree $ \lambda $ and weight $ \omega $. The value of $ \varrho_K \left( \lambda, \omega \right)$ turns out to be of fundamental importance in calculations of quantities which give information about the dynamics of \textit{NK}-Kauffman networks, as: {\it i}) The probability of extracting a $K$-Boolean function with irreducible-degree $ \lambda $, through the distribution $ \Pi\left( \omega \right) $ Eq.~(8). This quantity is an ingredient in the study of the stability of \textit{NK}-Kauffman networks as see in Appendix~A Eq.~(A2). {\it ii}) The study of the phase transition of the \textit{NK}-Kauffman networks done in Ref.~[10] (see Appendix~B for a summary).

We are using this framework further as a tool for the calculation of the dynamical properties of \textit{NK}-Kauffman networks with promising results.

\section{Acknowledgments}

This work is supported in part by {\bf PAPIIT} projects No.~{\bf IN101309-3} and {\bf IN102712-3}. The authors wish to thank: The referee for his/her accurate work, and specifically for calling our attention about canalizing Boolean functions, Thal\'\i a Figueras for careful reading of the man\-u\-script, and Pilar L\'opez Rico for accurate services on informatics. The second author (FZ) thanks Maximino Aldana, and Alberto Verjovsky for fruitful mathematical discussions.

\section{Appendix~A: About the Canalizing Boolean Functions}

For clarification purposes, we state now in the mathematical framework of this article the concept of a {\it canalizing} $K$-Boolean function, which was first introduced by Kauffman~${}^{3,12}$.

\leftline{{\bf Definition}:}

\begin{itemize}

\item[] A \ $ b_K: \mathbb{Z}_2^K \to \mathbb{Z}_2 $ Boolean function is said to be {\it canalizing} iff $ \exists $ $ i \in \left[ K \right] $, and $ \xi, \tau \in \mathbb{Z}_2 $ such that
$$
S_i = \xi \ \ \Rightarrow \ \ b_K \left( S_1, \dots, S_{i-1}, \xi, S_{i+1}, \dots, S_K \right) = \tau ,
$$
    $ \forall \ S_j \in \mathbb{Z}_2 $ with $ j \in \left[ K \right] \setminus \left\{ i \right\} $.

 \end{itemize}

\noindent
As examples for $ K = 2 $: in Table~{\it 1} all Boolean functions, but {\bf 7} and {\bf 10}, are canalizing. In more detail: function $ {\bf 15} = \vee $ also known as the {\bf OR} function, is canalizing since if one their two arguments ($ S_1 $ or $ S_2 $) is set to ``1", then $ {\bf OR} = 1 $ whichever the value of the other argument. We see that this definition divides $ \Xi_K $ in two disjoint subsets (the canalizing, and the no-canalizing sets).

{\it N.B.} Canalization and irreducibility are different properties of Boolean functions, neither of them containing the other. While canalizing functions are partially responsible of the frozen dynamics of \textit{NK}-Kauffman networks since they generate forcing structures that tend to trap the dynamics into short loops~${}^{3}$; they do not offer a full understanding of their dynamics. This requires a different and shifter information of the behavior of $ b_K $ as a function of their arguments; as irreducibility does~${}^{9}$. Irreducibility appears in a natural way in the calculation of important quantities that govern the dynamics of \textit{NK}-Kauffman networks~${}^{9-11,16}$. We quote three examples from our researches:

\begin{itemize}

\item[{\it i})] The calculation of the mean number of \textit{NK}-Kauffman networks mapped to the same functional graph $ \vartheta \left( N, K \right) $, was done in the Appendix of Ref.~[9] equations (R18), and (R19); with the result
$$
\vartheta \left( N, K \right) = \left\{ 1 - \varphi \left( N, K
\right)  \right\}^{-N}, \eqno(R18)
$$
where
$$
\varphi \left( N, K \right) = {\sum_{\lambda=0}^K \beta_K \left( \lambda \right) \left[ {N - \lambda \choose K - \lambda} - 1 \right] \over 2^{2^K} {N \choose K} }. \eqno(R19)
$$
As we see, this important quantity, is expressed as a series generated by $ \beta_K \left( \lambda \right) $ (the number of Boolean functions with irreducible-degree $ \lambda $). So, decomposition (11) with their cardinalities (15) contain the information of the dynamical diversity of \textit{NK}-Kauffman networks. Furthermore, each term of the series becomes smaller with growing $ \lambda $; which yields, for $ N \gg 1 $, and $ K \sim {\cal O} \left( \ln \ln N \right) $, the asymptotic expression~${}^{9,11}$
$$
\varphi \left( N, K \right) \approx {1 \over 2^{2^K - 1} } \left[ 1 + {\cal
O} \left( {1 \over N} \right) \right] .
$$
Using this result into (R18) and looking for a value $ K_c $ (of $ K $) such that $ \vartheta \left( N, K_c \right) = 1/2 $, it is obtained that a critical connectivity $ K_c $ exists, such that:

\begin{itemize}

\item[{\it a})] It is given by
$$
K_c \approx \log_2 \log_2 \left( {2 N \over \ln 2} \right) + {\cal
O} \left( {1 \over N \ln N } \right). \eqno(R23)
$$

\item[{\it b})] For $ K < K_c $ $ \vartheta \left( N, K \right) \gg 1 $, so many \textit{NK}-Kauffman networks are mapped into the same functional graph.

\item[{\it c})] For $ K > K_c $; $ \vartheta \left( N, K \right) \approx 1 $. So in this case, almost any \textit{NK}-Kauffman network is mapped to a different functional graph.

\end{itemize}

See Refs.~[9,11] for details.

\item[{\it ii})] The calculation for the probability $ P \left( {\cal A} \right) $; that an \textit{NK}-Kauffman network [Appendix~B Eq.~(B1)] remains invariant against a change in one of their $K$-connection functions $ C_K^{*(i)} $ was reported in Eqs.~(23) of Ref.~[9] (as one of the main ingredients for final result) obtaining,
$$
P \left( {\cal A} \right) = \sum_{\lambda = 0}^K {K! \left( N - \lambda \right)! \over N! \left( K - \lambda \right)!} \ P \left[ b_K \in \mathfrak{T}_K \left(\lambda \right) \right] , \eqno(A1)
$$
where $ P \left[ b_K \in \mathfrak{T}_K \left(\lambda \right) \right] $ is the probability to extract a function with irreducible-degree $ \lambda $. Once again we see other important quantity that gauges the dynamical behavior of \textit{NK}-Kauffman networks and is expressed by a series of quantities that are function of decomposition (11) of $ \Xi_K $ in their irreducible subsets $ \mathfrak{T}_K \left(\lambda \right) $.

Note that $ P \left[ b_K \in \mathfrak{T}_K \left(\lambda \right) \right] $ is analytically expressed by the use of (8) and (31) giving
$$
P \left[ b_K \in \mathfrak{T}_K \left(\lambda \right) \right] = \sum_{\omega=0}^{2^K} \varrho_K \left( \lambda, \omega \right) \ \Pi \left( \omega \right) . \eqno(A2)
$$
In Ref.~[9], however, the problem was focused in the asymptotic behavior of (A1) for $ N \gg 1 $, and $ K \sim {\cal O} \left( 1 \right) $, thus obtaining
$$
P \left( {\cal A} \right) \approx P \left[ b_K \in {\cal I}_K \left( 0 \right) \right] + {\cal O} \left( {1 \over N} \right) = p^{2^K} + \left( 1 - p \right)^{2^K} + {\cal O} \left( {1 \over N} \right)
$$
without requiring the full calculation of (A2).

\item[{\it iii})] The transition curve for the dynamics of \textit{NK}-Kauffman networks in a mean field treatment is shown to depend in the average connectivity, which is a function of $ \lambda \left( b_K \right) $ and not of canalization. See Appendix~B, where the transition curve (B2) is obtained as the average of $ \lambda \left( b_K \right) $ weighted by (8) and (31); see also Ref.~[10] for a more detailed study of the mean field treatment.

\end{itemize}

\newpage

\section{Appendix~B: Mean Field Dynamics of \textit{NK}-Kauffman Networks}

We summarize the mean field approach, corrected for irreducibility, to study the dynamics of the \textit{NK}-Kauffman networks. A detailed study should be consulted in Ref.~[10].

\textit{NK}-Kauffman networks, are Boolean endomorphisms $ f: \mathbb{Z}_2^N \longrightarrow \mathbb{Z}_2^N $ of the form
$$
\mathbb{Z}_2^N \ \stackrel{\ C_K^{*(i)}}{\longrightarrow} \ \mathbb{Z}_2^K \
\stackrel{\ b_K^{(i)}}{\longrightarrow} \ \mathbb{Z}_2 \hskip1.0cm i = 1, \dots, N ,
$$
where the {\it connection function} $ C_K^{*(i)} : \mathbb{Z}_2^N \rightarrow \mathbb{Z}_2^K $ cuts $ N - K $ of the $ N $ Boolean variables $ S_j $, $ j = 1, \dots, N $, so $ C_K^{*(i)} \left( S_1, \dots, S_N \right) = \left( S_{i_1}, \dots, S_{i_K} \right) $, with $ \left\{ i_1, \dots, i_K \right\} \subseteq \left[ N \right] $ ($ i_\alpha \in \left[ N \right] $, $ \alpha \in \left[ K \right] $) being whichever of the $ {N \choose K} $ subsets with cardinality $ K $ of $ \left[ N \right] $. The dynamic is defined by the synchronous iterations
$$
S_i \left( t + 1 \right) = b_K^{(i)} \circ C_K^{*(i)}
\left( {\bf S} \left( t \right) \right), \ \ i = 1, \dots, N . \eqno(B1)
$$
The $ b_K $ Boolean functions are extracted randomly according to the probability distribution (8), while the {\it connection functions} $ C_K^{*(i)} $ are extracted with equiprobability from the $ {N \choose K} $ possible ones.

The problem, is to study how is the way in which the dynamics of (B1) behaves as a function of their defining parameters, which are $ N $, $ K $, and the bias $ p $ of the probability distribution (8) with which the $ b_K $ functions are extracted.

A way to observe how the dynamics of (B1) evolves is to see the behavior of the {\it Hamming} distance of two nearby states $ {\bf S}, {\bf S'} \in \mathbb{Z}_2^N $
$$
d_H \left( {\bf S}, {\bf S'} \right) = \sum_{i=1}^N \left[ S_i + S'_i \right]_2
$$
for asymptotically big values of $ N $. The dynamic generated by (B1) is deterministic, but the construction of the functions $ b_K^{(i)} $ and $ C_K^{*(i)} $ which determines the endomorphism is done randomly. This allows to do a statistical treatment of the dynamics and make a {\it mean field approximation}~${}^{2,4,10}$. For that: let us see, that due to the randomness of the construction of (B1), each site $ i $ at $ t = 0 $, such that $ S_i (0)\neq  S'_i (0) $, will affect on average $ K $ sites; each one of them is going to be, also, the argument of a $ b_K $ function at the next iteration $ t = 1 $. So, at the next step, site $ i $ will contribute on average to the Hamming distance by the factor
$$
\Phi_i = \left< P_c ( b_K ) \ \lambda ( b_K ) \right>_i ,
$$
with the average taken with respect to the $ b_K^{(i)} $ that contribute to the  $i$-site, and $ P_c ( b_K ) $ being the probability that each $ b_K^{(i)} $ changes its output, due that one of their arguments has changed, (which is explicitly calculated in Ref.~[10]). If the system starts at an initial Hamming distance $ d_H (0) \equiv d_H \left( {\bf S} (0), {\bf S'} (0) \right) $ such that $ 1 \ll d_H (0) \ll N $, for $ N \gg 1 $ we may apply the central limit theorem and take the average over the $ N $-sites
$$
\left< \Phi \right> = { 1 \over N} \ \sum_{i=1}^N \Phi_i \equiv \left< P_c ( b_K ) \ \lambda ( b_K ) \right> .
$$

Now, at $ t = 1 $ Hamming distance will grow (or decay) on average by
$$
d_H (1) \approx d_H (0) \left< P_c ( b_K ) \ \lambda ( b_K ) \right> ,
$$
or more generally, while the condition $ 1 \ll d_H (t) \ll N $ is fulfilled
$$
d_H (t + 1) \approx d_H (t) \left< P_c ( b_K ) \ \lambda ( b_K ) \right> .
$$
Solving this difference equation we obtain
$$
d_H (t) \approx d_H (0) \left< P_c ( b_K ) \ \lambda ( b_K ) \right>^t .
$$
So depending on whether $ \left< P_c ( b_K ) \ \lambda ( b_K ) \right> $ is greater or lower than $ 1 $, Hamming distance will grow or decay exponentially; with the equation
$$ \Delta \left( K, p \right) \equiv \left< P_c ( b_K ) \ \lambda ( b_K ) \right> = 1
$$
signaling the phase transition frontier. $ \Delta \left( K, p \right) $ may be calculated explicitly through the use of (31) obtaining the equation
$$
\Delta \left( K, p \right) = \sum_{ \omega = 0}^{2^K} \, \Pi \left( \omega \right) \, P_c \left( \omega \right) \, \sum_{\lambda = 0}^K \ \lambda \ \varrho_K \left( \lambda, \omega \right) \eqno(B2)
$$
which is explicitly calculated in Ref.~[10], obtaining;

\newpage

$$
\Delta \left( K, p \right) =  2 \, K \, p \left( 1 - p \right) \times \hskip9cm
$$
$$
\hskip3.75cm \times \left\{ 1 - 2 \, p \, \left( 1 - p \right) \left[ 1 - 2 \, p \, \left( 1 - p \right) \right]^{2^{K-1}-2} \right\} = 1 , \eqno(B3)
$$
for the phase transition curve. Equation (B3) is an improvement of a result previously obtained in 1986 by Derrida \& Stauffer Ref.~[15], for the mean field treatment, where reducibility of Boolean functions was not taken into account; obtaining the result
$$
\Delta \left( K, p \right) =  2 \, K \, p \left( 1 - p \right) = 1
$$
for the transition curve. See Ref.~[10] for more details.

\bigskip

\

{\bf References}

\begin{itemize}

\item[${}^{1}$] Kruskal, M.D., {\it The Expected Number of
Components under a Random Mapping Function}. Am.~Math.~Monthly
{\bf 61} (1954) 392; Harris, B., {\it Probability Distributions Related to Random Mappings}. Ann. Math.~Stat. {\bf 31} (1960) 1045; Frank Harary, {\it Graph Theory}. Addison-Wesley (1972).

\item[${}^{2}$] Hertz,  J., Krogh, A., and Palmer,  R. G., Introduction to the Theory of Neural Computation (Addison-Wesley, Redwood
City, CA, 1991).

\item[${}^{3}$] Kauffman, S.A., {\it The Origins of Order: Self-Organization and Selection in Evolution}. Oxford University Press (1993).

\item[${}^{4}$] Aldana, M., Coppersmith, S. and Kadanoff, L., {\it Boolean Dynamics with Random Couplings}. In: Perspectives and Problems in Nonlinear Science, 23--89. Springer Verlag, New York (2003).

\item[${}^{5}$] Weisbuch, G., {\it Complex Systems Dynamics}. Addison Wesley, Redwood City, CA (1991); Wolfram, S., {\it Universality and Complexity in Cellular Automata}. Physica~D {\bf 10} (1984) 1.

\item[${}^{6}$] Kauffman, S.A., {\it Metabolic Stability and Epigenesis in Randomly Connected Nets}. J.~Theoret.~Biol. {\bf 22} (1969) 437.

\item[${}^{7}$] Derrida, B., and Flyvbjerg, H., {\it The Random Map Model: a Disordered Model with Deterministic Dynamics}. J.~Physique {\bf 48} (1987) 971; Romero, D., and Zertuche, F., {\it The Asymptotic Number of Attractors in the Random Map Model}. J.~Phys.~A:~Math.~Gen. {\bf 36} (2003) 3691; {\it Grasping the Connectivity of Random Functional Graphs}. Stud. Sci. Math. Hung. {\bf 42} (2005) 1.

\item[${}^{8}$] Flyvbjerg, H., and Kjaer, N.J., {\it Exact Solution of Kauffman's Model with Connectivity One}. J.~Phys.~A:~Math.~Gen. {\bf 21} (1988) 1695.

\item[${}^{9}$] Zertuche, F., {\it On the robustness of NK-Kauffman networks against changes in their connections and Boolean functions}. J.~Math.~Phys. {\bf 50} (2009) 043513.

\item[${}^{10}$] Zertuche, F., {\it Boolean Irreducibility and Phase Transitions in \textit{NK}-Kauffman Networks}. \texttt{Submitted for publication 2012}.

\item[${}^{11}$] Romero, D., and Zertuche, F., {\it Number of Different Binary Functions Generated by \textit{NK}-Kauffman Networks and the Emergence of Genetic Robustness}. J.~Math.~Phys. {\bf 48} (2007) 083506.

\item[${}^{12}$] Kauffman, S.A., {\it Gene Regulation Networks: A Theory for their Global Structure and Behavior}. Current Topics in Dev.~Biol. {\bf 6} (1971) 145; {\it The Large-Scale Structure and Dynamics of Gene Control Circuits: An Ensemble Approach}. J. Theoret.~Biol. {\bf 44} (1974) 167.

\item[${}^{13}$] Comtet, L., {\it Advanced Combinatorics}. Reidel, 1974, p. 165.

\item[${}^{14}$] Hausdorff, F., {\it Set Theory}. Chelsea Pub. Comp. 2nd Ed. (1957); R.~R.~Slotl, {\it Set Theory and Logic}. Dover (1979).

\item[${}^{15}$] Derrida, B., and Stauffer, D., {\it Phase Transitions in Two-Dimensional Kauffman Cellular Automata}. Europhys.~Lett. {\bf 2} (1986) 739.

\item[${}^{16}$] Zertuche, F., {\it Work in progress}.

\end{itemize}

\end{document}